\documentstyle[12pt]{article}
\begin{document}
\thispagestyle{empty}
\hspace*{\fill} PAR-LPTHE 96-12\\[.5cm]
\begin{center}{ \bf\sc\Large 
Coupling supergravity to non-supersymmetric matter }\\[.5cm]
{\sc J. A. Teschner\footnote{The author thanks the DFG for financial support}}\\[.5cm]
{Laboratoire de Physique Th\'{e}orique et Hautes Energies,\\
Universit\'{e} Pierre et Marie Curie, Paris VI,\\
Universit\'{e} Denis Diderot, Paris VII,\\
Bte 126, 4 Place Jussieu, 75252 Paris cedex 05, France\\
teschner@lpthe.jussieu.fr}\\[.5cm]
{March 1996}
\end{center}
\begin{abstract}
By introducing a nonlinearly transforming goldstino field non-super\-sym\-metric
matter can be coupled to supergravity. This implies the possibility of 
coupling a standard model with one Higgs to supergravity. 
\end{abstract}
\parindent0.3cm
\parskip0.5ex
\newenvironment{aufz}{\begin{list}{$\bullet$}{\topsep0ex \parsep0ex \itemsep0ex}}{\end{list}}
\renewcommand{\a}{\alpha}
\newcommand{\ad}{\dot{\alpha}}
\renewcommand{\b}{\beta}
\newcommand{\bd}{\dot{\beta}}
\renewcommand{\c}{\gamma}
\newcommand{\cd}{\dot{\gamma}}
\newcommand{\g}{{c_{+}}}
\newcommand{\gc}{\mbox{$\bar{c}_{+}$}}
\newcommand{\agc}{{\bar{c}_{-}}}
\newcommand{\ag}{{c_{-}}}
\newcommand{\la}{{\cal L}}
\newcommand{\ca}{c_{+}^{\alpha}}
\newcommand{\cca}{\bar{c}_{+\dot{\alpha}}}
\renewcommand{\aa}{c_{-}^{\alpha}}
\newcommand{\aca}{\bar{c}_{-}^{\dot{\alpha}}}
\newcommand{\cb}{c_{+}^{\beta}}               
\newcommand{\ccb}{\bar{c}_{+\dot{\beta}}}
\newcommand{\da}{{\cal D}_{\alpha}}
\newcommand{\dA}{{\cal D}_A}
\newcommand{\dB}{{\cal D}_B}
\newcommand{\dC}{{\cal D}_C}
\newcommand{\dac}{\bar{\cal D}^{\dot{\alpha}}}
\newcommand{\dd}{\delta^3(\vec{x}-\vec{x}')}
\newcommand{\ddk}{\delta^3(\vec{k}-\vec{k}')}
\newcommand{\ba}{b^{\alpha}}
\newcommand{\bac}{\bar{b}_{\dot{\alpha}}}
\newcommand{\bb}{b^{\beta}}               
\newcommand{\cbc}{\bar{b}_{\dot{\beta}}}
\newcommand{\f}{\varphi}
\newcommand{\fc}{\bar{\varphi}}
\newcommand{\gma}{{\psi_m^{\alpha}}}
\newcommand{\ga}{{\psi_a}}
\newcommand{\gad}{{\ga^{3/2}}}
\newcommand{\gac}{{\bar{\psi}_a}}
\newcommand{\gb}{{\psi_b}}
\newcommand{\gbc}{{\bar{\psi}_b}}
\newcommand{\x}{(\vec{x},t)}
\newcommand{\xs}{(\vec{x}',t)}
\newcommand{\s}{\sigma}
\renewcommand{\sc}{\bar{\sigma}}
\newcommand{\sA}{\sigma_a}
\newcommand{\sab}{\sigma^{ab}}
\newcommand{\sabc}{\bar{\sigma}^{ab}}
\newcommand{\sac}{\bar{\sigma}^a}
\newcommand{\sbc}{\bar{\sigma}^b}
\newcommand{\scc}{\bar{\sigma}^c}
\newcommand{\sa}{\sigma^a}
\renewcommand{\sb}{\sigma^b}
\renewcommand{\pm}{\partial_m}
\newcommand{\p}{\partial}
\newcommand{\sm}{\sigma_m}
\newcommand{\pM}{\partial^m}
\renewcommand{\pc}{{\bar{\psi}}}
\newcommand{\r}{\rho}
\newcommand{\rc}{\bar{\rho}}
\newcommand{\sM}{\sigma^m}
\renewcommand{\l}{\lambda}
\newcommand{\lc}{\bar{\lambda}}
\newcommand{\vm}{\!\!<\!\!M\!\!>\!\!}
\newcommand{\fr}[2]{{\textstyle\frac{#1}{#2}}}
\newcommand{\ed}{(e^{-i\vec{k}(\vec{x}-\vec{x}')}+e^{i\vec{k}(\vec{x}-\vec{x}')})}
\newcommand{\D}{{\cal D}}
\newcommand{\db}{{\cal D_{\cal B}}}
\newcommand{\dc}{{\cal D_{\cal C}}}
\newcommand{\Ph}{\hat{A}}
\renewcommand{\a}{\alpha}
\renewcommand{\b}{\beta}
\renewcommand{\c}{\gamma}
\newcommand{\al}{\tilde{\a}}
\newcommand{\bl}{\tilde{\b}}
\newcommand{\cl}{\tilde{\gamma}}
\renewcommand{\d}{{\cal D}}
\renewcommand{\sc}{\bar{\sigma}}
\renewcommand{\l}{\lambda}
\renewcommand{\pc}{\bar{\psi}}
\newcommand{\xc}{\bar{\xi}}
\renewcommand{\t}{\Theta}
\newcommand{\tc}{\bar{\Theta}}
\newcommand{\td}{{\Theta\cal D}}
\newcommand{\vh}{\hat{V}}
\newcommand{\vt}[1]{\tilde{V}_{Ai}^{(#1)}}
\section{Introduction} 
The original aim to study nonlinear realizations of symmetries was to investigate
the physical consequences of a symmetry if it is spontaneously broken. In that case the
symmetry is hidden in the effective low-energy theory. If the symmetry is
global, its effects can always be described in terms of an inhomogenously
transforming field, the Goldstone-field, and a characteristic structure of
its couplings \cite{cwz}. Similiar results have been obtained in the context of 
global supersymmetries in \cite{gssn} and \cite{uz}. In particular, the construction
of Lagrangians for nonlinearly realized global supersymmetries was discussed in \cite{uz}.

Effective low-energy theories of spontaneously
broken local symmetries also allow for a description in terms of Goldstone-fields,
but now it is an unphysical degree of freedom, so that its couplings are
unobservable \cite{ku}. In that case the symmetry manifests itself mainly via
gauge boson couplings, leading for example to effective four fermion interactions
at low energies.

The aim of this letter is to show that a similiar situation is to be found
for (d=4,N=1) supergravity theories. A construction is presented that allows to couple 
non-supersymmetric matter to supergravity by using a nonlinearly transforming 
goldstino field. The goldstino field is unphysical, so that the physical consequences of
supergravity come from the couplings of the matter-fermions to the gravitino and the
corresponding effective four fermion interactions at low energies. 
 
As an interesting consequence local supersymmetry
allows a standard model with only one Higgs boson coupled to
supergravity. 

Similiar results have recently independently been obtained by P. John (private communication).

The conventions are those of \cite{wb}, while the approach to supergravity
is that of \cite{dr}.
  
\section{Nonlinear transformation laws}
Let me first collect the results to be derived in this section:
\begin{aufz}
\item There exists a consistent nonlinear transformation law 
for a so-called Goldstino field,
which is of the following form:
\begin{eqnarray} \delta_{\xi}\l^{\a} & = & \eta_{\xi}^{\a}(\l,\bar{\l},E_A{}^M)-\eta_{\xi}^m(\l,\bar{\l},E_A{}^M)\p_m\l^{\a} \\
\eta_{\xi}^{\a} & = & \xi^{\a}+{\cal O}(\l).
\end{eqnarray}
Here $E_A^M$ symbolically denotes the fields of the supergravity multiplet.
\item 
Given a Lorentz- and gauge tensorfield $V_{Ai}$ one can define local
supersymmetry transformations by
\begin{equation} \delta_{\xi}V_A=-\eta_{\xi}^m(\l,\bar{\l},E_A{}^M)\p_mV^A+\eta_{\xi A}{}^B(\l,\bar{\l},E_A{}^M)V_B
+\eta_{\xi}^I(\l,\bar{\l},E_A{}^M)\delta_I V_A. \end{equation}
This transformation law 
realizes the algebra with the help of the Goldstino field and
will be called a standard matter transformation law. 

These two results were first derived in \cite{sw1} and \cite{sw2} by educated
guess. The basic idea for a systematic construction is also due to J. Wess,
and was developed in \cite{ra}. I will use a similiar but simpler version.
\item In a supersymmetry multiplet it is possible to replace some components
by composite fields build out of the remaining components and the Goldstino.
Even if something similiar was done in \cite{sw2}, the approach used below
seems to be new.
\end{aufz}

For the first step of this construction one 
defines the components of a multiplet by
\begin{equation} k=0\ldots4 : V^{(k)}_{Ai\al_k\ldots\al_1} = \fr{1}{k!}\D_{{[}\al_k}\ldots\D_{\al_1{]}}V_{Ai}.
\end{equation} 
It is important to note, that the spinor derivatives used here are assumed
to be gauge covariant, in contrast to \cite{wb}, \cite{sw1}, \cite{sw2}.
By introducing constant, anticommuting variables $\t^{\al}=(\t^{\a},\bar{\t}_{\ad})$,
it is possible to collect the component fields in a generating function: 
\begin{eqnarray}
\vh_{Ai} & = & \sum_{k=0}^4\t^{\al_1}\ldots\t^{\al_k}V^{(k)}_{Ai\al_k\ldots\al_1} \\
 & = & \sum_{k=0}^4\frac{1}{k!}(\t\D)^kV_{Ai}.
\end{eqnarray}
Note that the $\t$'s anticommute with all generators of the supergravity
algebra. Therefore $\vh_{Ai}$ transforms as follows:
\begin{equation} \delta_{\xi}\hat{V}_{Ai}=
-\sum_{k=0}^4\t^{\al_1}\ldots\t^{\al_k}\xi\D V^{(k)}_{Ai\al_k\ldots\al_1}.
\end{equation}
The crucial step now is the following:
It will be shown that one can find functions 
\begin{equation} \begin{array}{cccc}
\eta_{\al}{}^{\bl}(\t,E_A{}^M) & \eta_{\al}{}^{m}(\t,E_A{}^M) & 
\eta_{\al A}{}^B(\t,E_A{}^M) & \eta_{\al}{}^I(\t,E_A{}^M) \end{array}, \end{equation}
such that the transformation law takes the following form
($\p_{\al}=\frac{\p}{\p\t^{\al}}$):
\begin{eqnarray}
\D_{\al}\vh_{Ai}(x,\t) & = & \sum_{k=0}^4\t^{\al_1}\ldots\t^{\al_k}\D_{\al}V^{(k)}_{Ai\al_k\ldots\al_1} \\
 & = & (\eta_{\al}{}^m\p_m+\eta_{\al}{}^{\bl}\p_{\bl}+\eta_{\al}{}^I\delta_I)\vh_{Ai}-\eta_{\al A}{}^B\vh_{Bi}
\end{eqnarray}
These functions $\eta$ will be the building-blocks that appear in the
nonlinear transformation laws. The proof of this statement procedes in three
steps.
\begin{aufz}
\item The first step is to show that
\begin{equation} 
\D_{\al}\vh_{Ai}=\big(f_{\al}{}^b\D_b+f_{\al}{}^{\bl}\p_{\bl}+f_{\al}{}^I\delta_I+f_{\al}{}^{ab}l_{ab}\big)\vh_A
\label{ast} \end{equation}
In lowest order in $\t$ this is trivial: Denoting the k-th order of the expansion
of f in powers of $\t$ by $f^{(k)}$, one has
\begin{equation} \begin{array}{lccr} f^{(0)}_{\al}{}^{\bl}=\delta_{\al}{}^{\bl} & 
f^{(0)a}_{\al}=0 & f^{(0)ab}_{\al}=0 & f^{(0)I}_{\al}=0 \end{array}. \end{equation}
The proof procedes by induction in the power of $\t$ with the help of the following
identity:
\begin{eqnarray}
\D_{\al}\frac{(\td)^n}{n!}V_A & = & -\sum_{k=1}^{n}\frac{(ad_{\td})^k(\D_{\al})}{(k+1)!}
\frac{(\td)^{n-k}}{(n-k)!}V_A+\p_{\al}\frac{(\td)^{n+1}}{(n+1)!}V_A, 
\end{eqnarray}
with
\begin{equation} \begin{array}{cc}
ad_{\td}(\D_{\al})={[}\td,D_{\al}{]} & 
(ad_{\td})^k(\D_{\al})=\overbrace{{[}\td,{[}\ldots{[}\td}^{k\:Faktors\:\td},\D_{\al}{]}\ldots{]}
\end{array} \end{equation}
Now assume \ref{ast} to be shown to $n-1$th order. On the right-hand side of
above identity there only appear terms $\D_{\al}(\td)^l V_A$ with $l\leq n-1$. 
Inserting the $l$th order of (\ref{ast}) leads to an expression, which is of the
desired form (\ref{ast}). Thus, for (\ref{ast}) to be valid to all orders in $\t$,
the functions $f$ have to satisfy the following recursion relations:
(Notation: The expansion of $(ad_{\td})^k(\D_{\al})$ with respect to the basis
$\D_{\al},\D_a,l_{ab},\delta_I$ will be written as follows: 
\begin{eqnarray} (ad_{\td})^k(\D_{\al}) & = & {[}(ad_{\td})^k(\D_{\al}){]}^{\bl}\D_{\bl}
+{[}(ad_{\td})^k(\D_{\al}){]}^a\D_a \nonumber \\
 & & {[}(ad_{\td})^k(\D_{\al}){]}^{ab}l_{ab}+{[}(ad_{\td})^k(\D_{\al}){]}^I\delta_I 
\hspace{1cm}.)\nonumber \\ 
f_{\al}{}^{\bl} & = & -{\bigg[}\sum_{k=1}^n\frac{(ad_{\td})^k 
(\D_{\al})}{(k+1)!}{\bigg]}^{\cl}f_{\cl}{}^{\bl} \\
f_{\al}{}^{\cal I} & = & -{\bigg[}\sum_{k=1}^n\frac{(ad_{\td})^k 
(\D_{\al})}{(k+1)!}{\bigg]}^{\cl}f_{\cl}{}^{\cal I}
-{\bigg[}\sum_{k=1}^n\frac{(ad_{\td})^k (\D_{\al})}{(k+1)!}{\bigg]}^{\cal I}
\end{eqnarray}
In the last line the index ${\cal I}$ can take the values $b,ab,I$. 
\item Denoting the representation matrices of the Lorentz transformation
$f_{\al}{}^{ab}l_{ab}$ by $(F_{\al})_{\bl}{}^{\cl}$ and $(F_{\al})_A{}^B$ one
has
\begin{equation} f_{\al}^{ab}l_{ab}(\t\D)^kV_{Ai}=\t^{\bl}(F_{\al})_{\bl}{}^{\cl}\p_{\cl}(\t\D)^kV_{Ai}
-F_{\al A}{}^B(\t\D)^kV_{Bi} \end{equation}
Now the transformation law reads
\begin{equation}
\D_{\al}\vh_{Ai}=(g_{\al}{}^a\D_a+g_{\al}{}^{\bl}\p_{\bl}+g_{\al}{}^I\delta_I)\vh_{Ai}-g_{\al A}{}^B\vh_{Bi},
\label{ast2} \end{equation}
with $g_{\al}{}^{\bl}=f_{\al}{}^{\bl}+\t^{\cl}(F_{\al})_{\cl}{}^{\bl}$,
$g_{\al A}{}^B=F_{\al A}{}^B$, $g_{\al}{}^a=f_{\al}{}^a$, and $g_{\al}{}^I=f_{\al}{}^I$.
\item To express the covariant derivatives $\D_a$ by partial ones, one has
to use
\begin{equation} \D_a=e_a{}^m(\D_m-\fr{1}{2}\psi_m{}^{\al}\D_{\al}), \end{equation}
with
\begin{equation} \D_m\vh_{Ai}=\p_m\vh_{Ai}+A_m{}^I\delta_I\vh_{Ai}-\t^{\al}\omega_{m\al}{}^{\bl}\p_{\bl}\vh_{Ai}-\omega_{mA}{}^B\vh_{Bi}. \end{equation}
This leads to the occurrence of spinor derivatives on the right hand side
again, but recursively inserting (\ref{ast2}) stops at some power of $\t$,
because $g_{\al}{}^a={\cal O}(\t)$. The functions $\eta$ are now defined by
the following recursion relations:
\begin{equation} \begin{array}{cc}
\eta_{\al}{}^m=g_{\al}{}^a\hat{g}_a{}^m & \hat{g}_a{}^m=e_a{}^m-\fr{1}{2}e_a{}^n
\psi_n{}^{\al}g_{\al}{}^b\hat{g}_b{}^m \\
\eta_{\al}{}^{\bl}=g_{\al}{}^{\bl}+g_{\al}{}^a\hat{g}_a{}^{\bl}-\eta_{\al}{}^m\t^{\cl}\omega_{m\cl}{}^{\bl} & 
\hat{g}_a{}^{\bl}=-\fr{1}{2}e_a{}^m\psi_m{}^{\al}(g_{\al}{}^{\bl}+g_{\bl}{}^b\hat{g}_b{}^{\bl}) \\
\eta_{\al A}{}^B=g_{\al A}{}^B+\eta_{\al}{}^m\omega_{mA}{}^B+g_{\al}{}^c\hat{g}_{cA}{}^B &
\hat{g}_{cA}{}^B=-\fr{1}{2}e_c{}^m\psi_m{}^{\al}(g_{\al A}{}^B+g_{\al}{}^a\hat{g}_{aA}{}^B) \\
\eta_{\al}{}^I=g_{\al}{}^I+\eta_{\al}{}^m A_m{}^I+g_{\al}{}^c\hat{g}_{c}{}^I &
\hat{g}_c{}^I=-\fr{1}{2}e_c{}^m\psi_m{}^{\al}(g_{\al}{}^I+g_{\al}{}^a\hat{g}_a{}^I)
\end{array} \end{equation}
\end{aufz}
Taking general coordinate transformations into account, the transformation law
$\delta_{\xi}\hat{V}_{Ai}=-(\xi^c\D_c+\xi^{\cl}\D_{\cl})\hat{V}_{Ai}$
takes the following form:
\[ \delta_{\xi}\hat{V}_{Ai}=-(\eta_{\xi}{}^{\al}\p_{\al}+\eta_{\xi}{}^m\p_m
+\eta_{\xi}{}^I\delta_I)\hat{V}_{Ai}+\eta_{\xi A}{}^B\hat{V}_{Bi} \]
From now on 
$\eta_{\xi}{}^{\cal I}=\xi^{\bl}\eta_{\bl}{}^{\cal I}+\xi^b\hat{g}_b{}^{\cal I}$,
${\cal I}\in \{ \al,m,{}_A{}^B,I \}$.

Now I am in the position to define nonlinear transformation laws
for the Goldstino ($\eta_{\xi}=\xi^{\al}\eta_{\al}$)
\begin{eqnarray} \delta_{\xi}\l^{\a}=-\xi^{\al}\D_{\al}\l^{\a} & = & \eta_{\xi}^{\a}(\l^{\a},\bar{\l}_{\ad},E_A{}^M)
-\eta_{\xi}^m(\l^{\a},\bar{\l}_{\ad},E_A{}^M)\p_m\l^{\a} \\
 & = & \xi^{\a}+{\cal O}(\l), \end{eqnarray}
and for standard matter fields $\delta_{\xi}\tilde{V}_{Ai}=-\xi^{\al}\D_{\al}\tilde{V}_{Ai}$
\begin{equation} =-\eta_{\xi}^m(\l^{\al},E_A{}^M)\p_m\tilde{V}_{Ai}
-\eta_{\xi}^I(\l^{\al},E_A{}^M)\delta_I\tilde{V}_{Ai}+\eta_{\xi A}{}^B(\l^{\al},E_A{}^M)\tilde{V}_{Bi}. \end{equation}
Note that these transformations can be equivalently written in the form
\begin{eqnarray}
\D_{\al}\l^{\b} & =  & g_{\al}{}^{\b}(\l^{\cl},E_A{}^M)-g_{\al}{}^b(\l^{\cl},E_A{}^M)\D_b\l^{\b} \\
\D_{\al}\tilde{V}_{Ai} & = & g_{\al}{}^b(\l^{\al},E_A{}^M) \D_b\tilde{V}_{Ai}-
  g_{\al A}{}^B(\l^{\al},E_A{}^M)\tilde{V}_{bi} +g_{\al}{}^I\delta_I\tilde{V}_{Ai}, 
\label{con} \end{eqnarray}
the functions $g$ being those defined in (\ref{ast2}).

To check the algebra one should note that by expressing the algebra of 
infinitesimal supersymmetry transformations on $\vh_{Ai}$ in terms of the
funktions $\eta$, one finds consistency conditions for them, namely
\begin{eqnarray*}
(\delta^{}_{{[}\zeta}\eta^m_{\xi{]}})
+(\eta^{\cal I}_{{[}\zeta}\p_{\cal I}\eta_{\xi{]}}^m) & = & \eta^m_{\xi^A\zeta^BT_{BA}} \\
(\delta^{}_{{[}\zeta}\eta^{\a}_{\xi{]}})
+(\eta^{\cal I}_{{[}\zeta}\p_{\cal I}\eta_{\xi{]}}^{\a}) & = &  
\eta^{\a}_{\xi^A\zeta^BT_{BA}}+\xi^A\zeta^BR_{BA\b}{}^{\a}\t^{\b}
\end{eqnarray*}
Similiar equations hold for $\eta_{\xi A}{}^B$ and $\eta_{\xi}{}^I$.

For the calculation of the algebra on $\l$ I need some useful notation:
$\delta^{(E)},\p^{(E)}_m$ only act on the supergravity multiplet, but not on
$\l$, while $\delta^{(\l)},\p^{(\l)}_m$ only act on $\l$; also
$\p_{\a}^{(\l)}=\frac{\p}{\p\l^{\a}}$.
\begin{eqnarray*}
{[}\delta_{\zeta},\delta_{\xi}{]}\l^{\a} & = &
\delta^{(E)}_{{[}\zeta}\eta^{\a}_{\xi{]}}+\eta_{{[}\zeta}^{\b}\p_{\b}^{(\l)}\eta_{\xi{]}}^{\a}
-\eta^m_{{[}\zeta}\p^{(\l)}_m\eta_{\xi{]}}^{\a} \\
 & & -\big(\delta^{(E)}_{{[}\zeta}\eta^m_{\xi{]}}+\eta^{\b}_{{[}\zeta}\p_{\b}^{(\l)}\eta^m_{\xi{]}}
-\eta^n_{{[}\zeta}\p^{(\l)}_n\eta^m_{\xi{]}}\big)\p_m\l^{\a} \\
 & & -\eta^m_{{[}\xi}\p_m\eta^{\a}_{\zeta{]}}+(\eta_{{[}\xi}^m\p_m\eta^n_{\zeta{]}})\p_n\l^{\a} \\
 & = & \delta^{(E)}_{{[}\zeta}\eta^{\a}_{\xi{]}}+\eta_{{[}\zeta}^{\b}\p_{\b}^{(\l)}\eta^{\a}_{\xi{]}}
+\eta^m_{{[}\zeta}\p_m^{(E)}\eta^{\a}_{\xi{]}} \\
 & & -\big(\delta^{(E)}_{{[}\zeta}\eta^m_{\xi{]}}+\eta^{\b}_{{[}\zeta}\p_{\b}^{(\l)}\eta^m_{\xi{]}}
+\eta^n_{{[}\zeta}\p_n^{(E)}\eta^m_{\xi{]}}\big)\p_m\l^{\a} \\
 & = & \eta^{\a}_{\xi^A\zeta^BT_{BA}}+\xi^A\zeta^B R_{BA\b}{}^{\a}\l^{\b})
-\eta^m_{\xi^A\zeta^BT_{BA}}\p_m\l^{\a} \\
 & = & \big(\delta_{\xi^A\zeta^BT_{BA}}+\fr{1}{2}\xi^A\zeta^BR_{BA}{}^{bc}l_{bc} \big)\l^{\a}
\end{eqnarray*}
The check of the algebra for the standard matter transformation law is similiar.
Because the description of the transformation law in terms of the functions
$\eta$ holds for real superfields, $\eta_{\xi}^m$,
$\eta_{\xi a}{}^b$ and $\eta_{\xi}^I$ are real, as well as $(\eta^{\a})^{\ast}=\eta^{\ad}$
and $\eta_{\a}{}^{\b}=-(\eta_{\ad}{}^{\bd})^{\ast}$. Therefore
the transformation law is consistent with $\l$ being
a Majorana spinor.

It is perhaps intersting to note that the transformation law for $\l$, although
looking somewhat different to that of \cite{sw1}, is not genuinely new:
It is possible to construct a function of a goldstino transforming as above
and the fields of the supergravity multiplet
that has the transformation law given in \cite{sw1}: see \cite{te}.

Finally I have to explain, how to define multiplets with components being
functions of $\l$ and a subset of independent component fields.
The key step is to define standard matter fields out of multiplets:
\begin{eqnarray*} \tilde{V}^{(0)}_{Ai} & = & \sum_{k=0}^4\frac{1}{k!}\l^{\al_1}(x)\ldots\l^{\al_k}(x)(\D_{\al_k}\ldots\D_{\al_1}V_{Ai})(x) \\
 & = & \vh_{Ai}(x,\l^{\al}(x))\:=\: V_{Ai}(x)+{\cal O}(\l) \\
\tilde{V}^{(k)}_{Ai\bl_k\ldots\bl_1}(x) & = & \sum_{k=0}^4\frac{1}{k!}\l^{\al_1}(x)\ldots\l^{\al_k}(x)
(\D_{\al_k}\ldots\D_{\al_1}\D_{{[}\bl_k}\ldots\D_{\bl_1{]}}V_{Ai})(x) \\
 & = & \vh^{(k)}_{Ai\bl_k\ldots\bl_1}(x,\l^{\al}(x))\:=\: V^{(k)}_{Ai\bl_k\ldots\bl_1}(x)+{\cal O}(\l)
\end{eqnarray*}
For notational simplicity I will only demonstrate, that $\tilde{V}^{0}_{Ai}$
transforms as standard matter field:
\begin{eqnarray*}
\lefteqn{\delta_{\xi}\tilde{V}^{(0)}_{Ai}(x^m) = 
\delta_{\xi}\vh_{Ai}(x^m,\l^{\al}(x^n))} \\
 & = & (\eta_{\xi}^{\al}-\eta_{\xi}^m\p_m\l^{\al})\frac{\p}{\p\l^{\a}}\vh_{Ai}
-\Big(\eta_{\xi}^{\al}\frac{\p}{\p\l^{\al}}+\eta_{\xi}^m\frac{\p}{\p x^m}\Big)\vh_{Ai}
+\eta_{\xi A}{}^B\vh_{Bi}-\eta_{\xi}^I\delta_I\vh_{Ai} \\
 & = & -\eta_{\xi}^m\frac{d}{dx^m}\vh_{Ai}(x^m,\l^{\al}(x^n))+\eta_{\xi A}{}^B\vh_{Bi}(x^m,\l^{\al}(x^n))
 -\eta_{\xi}^I\delta_I\vh_{Ai}(x^m,\l^{\al}(x^n)) \\
 & = & -\eta_{\xi}^m\p_m\tilde{V}^{(0)}_{Ai}(x^m)+\eta_{\xi A}{}^B\tilde{V}^{(0)}_{Bi}(x^m)
 -\eta_{\xi}^I\delta_I\tilde{V}^{(0)}_{Ai}(x^m) 
\end{eqnarray*}
Because $\delta \tilde{V}$ is proportional to $\tilde{V}$ it is
consistent to define supercovariant constraints by
demanding $\tilde{V}^{(k)}_{Ai\bl_k\ldots\bl_1}=0$ for some appropriate $k$. 
One can convince oneself
that these constraints can be solved to express the corresponding component
fields as functions of the goldstino and the remaining component fields:
To this aim write the definition of the $\tilde{V}^{(k)}_{Ai\bl_k\ldots\bl_1}$
in the form
\begin{equation} 
 V_{Ai}^{(k)}(x)=\vt{k}-\l^{\al}G_{A\al}^{(k)}{[}V_{Ai}^{(l)}{]} \label{rek}
\end{equation}
These equations are solved by recursively replacing $V^{(k)}_{Ai\bl_k\ldots\bl_1}$
by the right hand side of (\ref{rek}). Start with the lowest component $V^{(0)}_{Ai}$.
The order in $\l$ of the terms not depending on $\tilde{V}^{(l)}_{Ai}$ on the
right hand side must not increase every step of the iteration, because there
occur spatial derivatives of $V^{(0)}_{Ai}$, leading in the 
next iteration steps to derivatives of $\l$. But at least every second iteration
step increases the order of $\l$, because every spatial derivative only occurs
multiplied with at least two factors of $\l$ (basically because of 
${[}\D_{\al},\D_{\bl}{]}=\ldots\D_{a}\ldots$). Therefore the iteration stops
after finitely many steps. Solving for the next highest component of $V_{Ai}$, one
can now use the expression of $V^{(0)}_{Ai}$ in terms of the $\tilde{V}^{(k)}_{Ai}$,
and proceed accordingly to the case of $V^{0}_{Ai}$.
Obviously there is no problem to use this procedure also in the case of
constrained multiplets, such as chiral ones.
\section{Supersymmetrisation of non supersymmetric lagrangians}
Consider a lagrangian of the following form:
\begin{equation} L=e(L_{gravity}+L_{matter}) \end{equation}
Let it depend on
scalar fields $h_i$ and spinor fields $q_{Fx\a}$, being tensors with
respect to some internal gauge group, with kinetic terms 
\begin{equation} \begin{array}{cc} L_{kin,h}=-\frac{1}{2}g^{mn}D_m\bar{h} D_n h &
L_{kin,q}=-i\bar{q}\bar{\s}^a e_a{}^m D_mq \\
D_m h_i=\p_m h_i-A_m{}^I(T_I)_i{}^j h_j &
D_m q_{x\a}=\p_m q_{x\a}-\omega_{m\a}{}^{\b}q_{x\b}-A_m{}^I(T_I)_x{}^y q_{y\a},
\end{array} \end{equation}
and the gauge connections $A_m{}^I$ with kinetic term
\begin{equation} \begin{array}{cc}
\multicolumn{2}{c}{L_{kin,A}=-\frac{1}{4}g^{mn}g^{kl}trF_{mk}F_{nl}} \\
F_{mk}=F_{mk}{}^I\delta_I & F_{mk}{}^I=\p_m A_k{}^I-\p_k A_m{}^I
-c_{KJ}{}^I A_m{}^K A_k{}^J. 
\end{array} \end{equation}
Assume $L_{matter}$ to be of the form $L_{matter}=L_{kin}+L_{Yuk}+V(h^i)$,
where 
\begin{equation} L_{Yuk}=K^{FF'}h^x q_Fq_{F'x}+h.c. \end{equation}
I have introduced flavorindices $F,F'$, and $K^{FF'}$ denotes the matrix
of coupling constants.
Further, let the potential have the following properties:
($<{\cal O}>$ denotes the vacuum expectation value of the operator in brackets)
\begin{equation} \begin{array}{cc}
<\!V(h_i)\!>=V(0)=0 & <\!\frac{\p V}{\p h_i}(h_i)\!>=\frac{\p V}{\p h_i}(0)=0
\end{array} \end{equation}

The first step in the supercovariantization consists in the introduction of
appropriate superfields corresponding to the field variables of the above
lagrangian. In the case of the spinor and scalar fields one just takes the
corresponding standard matter fields $\tilde{h}$ and
$\tilde{q}$ ($\D_{\al}\tilde{h}={\cal O}(\l)$ and
$\D_{\al}\tilde{q}={\cal O}(\l)$).
Regarding the vector fields one has to define a field strength superfield,
in which the components $W_{\a}{}^I$ and $D^I$ are no independent fields
but functions of $\l,E_A{}^M$ and the field strength $F^I_{ab}$.
That this is indeed possible was explained in the preceeding section:
The appropriate constraints are $\tilde{W}^I_{\a}=0$ and $\tilde{D}^I=0$.
Remember $W_{\a}{}^I(x)={\cal O}(\l)$.

The next step is to define a superfield $\hat{L}_{matter}$ out of $L_{matter}$
by the following replacements:
\begin{eqnarray*}
L_{kin,h} & \rightarrow & \tilde{L}_{kin,h}=-\fr{1}{2}\hat{\D}_a\bar{\tilde{h}}\hat{\D}^a\tilde{h} \\
L_{kin,q} & \rightarrow & \tilde{L}_{kin,q}=-i\bar{\tilde{q}}\bar{\s}^a\hat{\D}_a \tilde{q} \\
L_{kin,A} & \rightarrow & \tilde{L}_{kin,A}=-\fr{1}{4}tr\hat{F}_{ab}\hat{F}^{ab} \\
L_{Yuk} & \rightarrow & \tilde{L}_{Yuk}=K^{FF'}\tilde{h}^x\tilde{q}_F\tilde{q}_{F'x} \\
V(h_i) & \rightarrow & \tilde{V}(\tilde{h_i})
\end{eqnarray*}
Note that $F^I_{ab}$ has been replaced by the supercovariant field strength
(see i.e. \cite{dr},\cite{wb}).
\begin{equation} \begin{array}{lr}
\multicolumn{2}{r}{ F_{ab}{}^I=e_a{}^m e_b{}^n(\p_m A_n{}^I-\p_n A_m{}^I
-c_{JK}{}^I A_m{}^J A_n{}^K)} \\
 & +\frac{i}{2}\bar{W}\bar{\s}_{{[}b}\psi_{a{]}}-\frac{i}{2}\bar{\psi}_{{[}a}\bar{\s}_{b{]}}W.
\end{array} \end{equation}
$\tilde{L}_{matter}=\tilde{L}_{kin}+\tilde{L}_{Yuk}+\tilde{V}$
is a gauge {\it invariant} and super{\it covariant} field.

The final step now is to define the manifestly supersymmetric lagrangian
\begin{equation} {\cal L}={\cal E}\Big(\frac{M}{2\kappa^2}+f-
\frac{3k^2}{8}(\bar{\D}^2+\fr{4}{3}M)\l^2\bar{\l}^2(1+\frac{2k^2}{6}\tilde{L}_{Materie})\Big)+h.c.,
\end{equation}
where
${\cal E}(\ldots)$ is the density formula of ref. \cite{dr}.
$\lambda_{\a}$ denotes the goldstino superfield, with
constraints
\begin{eqnarray}
\D_{\a}\l^{\b} & = & \fr{1}{k}\delta_{\a}{}^{\b}+\fr{1}{3}M^{\ast}\l^2\delta_{\a}{}^{\b}
-\fr{1}{9}M\bar{\l}^2\delta_{\a}{}{\b} \nonumber \\
 & & -\fr{1}{12}\big(\l b\bar{\l}\delta_{\a}{}^{\b}-\fr{1}{3}\l_{\a}(b\bar{\l})^{\b}
+\l^{\b}(b\bar{\l})_{\a}\big) \nonumber \\
 & & +i(\sigma^a\bar{\l})_{\a}\D_a\l^{\b}+{\cal O}(\l^3) \\
\bar{\D}_{\ad}\l^{\b} & = & \fr{1}{36}\l^2b^{\b}{}_{\ad}-\fr{1}{9}\l^{\b}
\bar{\l}_{\ad}M \nonumber \\
 & & -i(\l\sigma^a)_{\ad}\D_a\l^{\b}+{\cal O}(\l^3) 
\end{eqnarray}
calculated from (\ref{con}) using the solutions of the Bianchi-identities
for minimal constraints given in \cite{wb} and the replacement
$\lambda\rightarrow -k\lambda$ (${[}1/k{]}=2$). $\l$ is thus of 
dimension 3/2. 
The parameter $f$ will be fixed below by requiring the vanishing of the cosmological constant.
The component expansion of the first two terms (with $M$ and $f$), is to be
found in \cite{wb} or \cite{dr}, that of the $\l$ terms has to be calculated
using the spinor derivatives of $\l$ given by the constraints. Regarding
the $\tilde{L}$ terms observe that a spinor derivative acting on one factor
of $\l$ gives a constant up to terms of higher order in $\l$. Therefore terms
without factors of $\l$ only emerge if all four spinor derivatives act on
one factor of $\l$ respectively. The terms not stemming from 
$\tilde{L}_{matter}$ are
\begin{eqnarray}
{\cal L} & = & e\Big[ \fr{1}{\kappa^2}\Big(-\fr{1}{2}{\cal R}-\fr{1}{3}M^{\ast}M+\fr{1}{3}b^ab_a
               +\fr{1}{2}\epsilon^{klmn}(\bar{\psi}_k\sc_l D_m\psi_n-
               \psi_k\s_l D_m\bar{\psi}_n)\Big) \nonumber \\
         &   & -f(M+M^{\ast}+\pc\sc^{ab}\pc_b+\psi_a\s^{ab}\psi_b) \nonumber \\
         &   & -\fr{3}{k^2}-\fr{3}{2}i\l\s^m\p_m\bar{\l}
               -\fr{3}{2}i\bar{\l}\sc^m\p_m\l+2M\bar{\l}^2+2M^{\ast}\l^2
               +\fr{9i}{4k}(\l\s^a\pc_a-\psi_a\s^a\bar{\l}) \nonumber \\
& & +{\cal O}(\l b\bar{\l})+{\cal O}(\l^3)\Big].
\end{eqnarray} 
${\cal R}$ is the curvature scalar.

The terms from $\tilde{L}_{matter}$ of lowest order in $\l$ will be collected
in $\hat{L}_{matter}$. They are
\begin{aufz}
\item $\hat{L}_{kin,h}=-\frac{1}{2}g^{mn}D_m\bar{h}\D_n h$ (because of 
$-\frac{1}{2}\psi_a{}^{\a}\D_{\a}h={\cal O}(\l)$)
\item $\hat{L}_{kin,q}=-i(\bar{q}^x\bar{\s}^a)^{\a}e_a{}^m(\p_m q_{x\a}
-\omega_{m\a}{}^{\b}q_{x\b}-A_m{}^I(T_I)_x{}^y q_{y\a})$. The spin connection
in terms of elementary fields reads
\begin{eqnarray*} \omega_{abc} & = & \fr{1}{2}(\Omega_{abc}-\Omega_{bca}
+\Omega_{cab}) \\
\Omega_{abc} & = & e_{nc}e_a{}^m\p_me_b{}^n+\fr{i}{2}\psi_a\sigma_c\bar{\psi}_b+(a\leftrightarrow b)
\end{eqnarray*}
\item $\hat{L}_{kin,A}=-\frac{1}{4}g^{mn}g^{kl}trF_{mk}F_{nl}$. Because of
$W_{\a}{}^I={\cal O}(\l)$ no gravitino couplings occur.
\item $\hat{V}=V,\hat{L}_{Yuk}=L_{Yuk}$.
\end{aufz}
The equations of motion for $b^a$ and $M$ read:
\begin{eqnarray}
M & = & -3\kappa^2 f+6\kappa^2\lc^2+{\cal O}(\l^3)+{\cal O}^{(3)}(\l) \label{be1} \\
b^a & = & {\cal O}(\l^2)+{\cal O}^{(3)}(\l) \label{be2}
\end{eqnarray}
(${\cal O}^{(3)}(\l)$ denotes trilinear terms of order 
${\cal O}(\l)$.)
Inserting (\ref{be1}),(\ref{be2}) in ${\cal L}$ leads to
\begin{eqnarray}
{\cal L} & = & e\Big(-\fr{1}{2}{\cal R}+3\kappa^2f-\fr{3}{k^2} \nonumber \\
         &   & -\fr{3}{2}i\l\s^m\p_m\lc-\fr{3}{2}i\lc\sc^m\p_m\l-6\kappa^2f(\l^2+\lc^2)
               +\fr{9i}{4k}(\l\s^a\pc_a-\psi_a\s^a\lc) \nonumber \\
         &   & \fr{1}{2}\epsilon^{klmn}(\pc_k\sc_lD_m\psi_n-\psi_k\s_l D_m\psi_n)
               -f(\pc_a\sc^{ab}\pc_b+\psi_a\s^{ab}\psi_b) \nonumber \\
         &   & +\hat{L}_{matter}+{\cal O}^{(3)}(\l)
               +{\cal O}(\l^3)\Big)            
\end{eqnarray}

To have vanishing cosmological constant, one has to choose
\begin{eqnarray*} f & = & \frac{1}{k^2\kappa^2} \end{eqnarray*}
Local supersymmetry is spontaneously broken: In a Lorentz invariant
ground state one must have $<\l_{\a}>=0$, so it cannot be invariant if $\l$
transforms inhomogenously.
Besides the particles described by $L_{matter}$ the particle spectrum
contains a graviton, a gravitino and a goldstino, the latter two being massive
by the supersymmetric Higgs effect. Because of the gauge freedom, not all
degrees of freedom are physical. On the classical level it is clear, that one
can fix a gauge by demanding $\l$ to vanish. The conclusion that $\l$ is
unphysical can also be reached by BRS quantization of the linearized theory
(\cite{te}).
\section{Remarks and open questions}
The point of the present paper is that local supersymmetry does not constrain the matter 
spectrum of the theory if one allows it to be nonlinearly realized. 
The physical consequences of supergravity come from the
couplings of the physical gravitino components to the fermionic matter fields
which are due to the fermionic contributions to the spin connection.
 
As an amusing aside note that
it is also possible to replace the gravitino by a composite field out of
the goldstino and the remaining supergravity multiplet: The necessary 
relation is most compactly written as 
\begin{equation} \D_a\l_{\a}=0 \end{equation}
With the help of $\D_a=e_a{}^m(\D_m-\fr{1}{2}\psi_m{}^{\al}\D_{\al})$ it is
easy to convince oneself that it can be solved to give a composite gravitino.
In that case, by fixing a gauge $\l=0$, the whole theory reduces to a usual
theory with matter coupled to gravitation. 

Any spontaneously broken supergravity model with linearly realized
supersymmetry can be rewritten in terms of goldstino and standard matter fields if a
field aquires a vacuum expectation value (see \cite{sw2}). It would be interesting to
know whether any lagrangian constructed in terms of goldstino and standard matter fields can
be viewed as an effective lagrangian of some model with
linearly realized, but spontaneously broken local supersymmetry,
when the contributions of
heavy fields can be neglected. Methods to construct
extensions with linearly realized symmetries out of nonlinear ones would be
very interesting to explore which high-energy physics is compatible with
known low-energy phenomenology. 

{\bf Acknowledgements}

The results presented in this letter have been obtained in my
Hannover University Diploma thesis \cite{te}. The author thanks 
Prof. W. Buchm\"{u}ller for suggesting the subject as well as helpful discussions and
encouragement. Thanks also to Prof. N. Dragon for valuable discussions.

\end{document}